\documentclass{pnastwo}
\pdfoutput=1
\usepackage{graphicx}
\usepackage{times}
\usepackage{amssymb,amsfonts,amsmath}
\begin{document}
\conflictofinterest{Conflict of interest footnote placeholder}
\track{Insert 'This paper was submitted directly to the 
PNAS office.' when applicable.}
\footcomment{Abbreviations: RVB, resonating valence bond; 
 dHvA, de Haas-van Alphen effect; SdH, Shubnikov-de Haas effect; DDW, $d$-density wave; SDW, spin density wave; DSC, $d$-wave superconductor; Y123, $\mathrm{YBa_{2}Cu_{3}O_{y}}$; Y124, $\mathrm{YBa_{2}Cu_{4}O_{8}}$: BZ, Brillouin zone; RBZ, reduced Brillouin zone; CuO, Copper oxide. }
\title{Fermi pockets and quantum oscillations of the Hall coefficient in high temperature superconductors}
\author{Sudip Chakravarty\thanks{To whom correspondence should be addressed. E-mail:
 sudip@physics.ucla.edu}
\affil{1}{Department of Physics and Astronomy, University of
California Los Angeles, Los Angeles, CA 90095, USA}
\and
Hae-Young Kee
\affil{2}{Department of Physics, University of
Toronto, Toronto, Ontario M5S 1A7, Canada}}

\maketitle
\begin{article}

\begin{abstract}
Recent quantum oscillation measurements in high temperature superconductors  in high magnetic fields and low temperatures have ushered in a new era. These experiments  explore the normal state  from which superconductivity arises and provide  evidence of a reconstructed  Fermi surface consisting of electron and hole pockets in a regime in which such a possibility was previously considered to be remote. More specifically, the Hall coefficient  has been found to oscillate according to the Onsager quantization condition, involving only fundamental constants and the areas of the pockets,  but  with a sign that is negative. Here we explain the observations with the theory that  the alleged  normal state exhibits a hidden order, the $d$-density wave, which breaks symmetries signifying time reversal, translation by a lattice spacing, and a rotation by an angle $\pi/2$, while the product of any two symmetry operations is preserved.  The success of our analysis  underscores  the importance of spontaneous  breaking of symmetries, Fermi surface reconstruction, and conventional quasiparticles. We primarily focus on the version of the  order that is commensurate with the underlying crystalline lattice, but also touch upon the consequences if the order were to incommensurate. It is shown that while commensurate order results in  two independent oscillation frequencies as a function of the inverse of the applied magnetic field, incommensurate order leads to  three independent frequencies. The oscillation amplitudes, however,  are determined by  the mobilities of the charge carriers comprising the Fermi pockets.
\end{abstract}

Any prospect of elucidating the mechanism of high temperature superconductivity in cuprates is remote without answering some of the basic questions in clear terms. The most important of which is the notion of a Fermi surface---whether or not  it exists, or it is reconstructed due to a broken symmetry in the pseudogap state~\cite{Chakravarty:2008}. An equally basic question is the extent to which a featureless spin liquid ground state, the resonating valence bond state (RVB),  is important~\cite{Anderson:1987}.    In this respect the recent experiments on quantum oscillations in high magnetic fields in high quality samples of $\mathrm{YBa_{2}Cu_{3}O_{y}}$ (Y123) and $\mathrm{YBa_{2}Cu_{4}O_{8}}$ (Y124)  have been striking~\cite{Doiron-Leyraud:2007,Bangura:2008,LeBoeuf:2007,Jaudet:2007,Yelland:2008}. 

The  quantum oscillations in the magnetization (de Haas van Alphen effect-dHvA), in the conductivity (Shubnikov-de Haas effect-SdH), and in the Hall coefficient ($R_{H}$) have  long been used to map out the Fermi surface and its topology in metals and semimetals~\cite{Shoenberg:1984}. A  Fermi surface  differentiates the occupied electronic states from the unoccupied states in the momentum space and plays a fundamental role in quantum theory of matter. The highest occupied energy is an important parameter called the Fermi energy. The excitations from this surface determine the quasiparticles, which behave in many ways like bare particles but their properties are modified by the collective interactions. An important notion is that a Fermi surface  is  a topological invariant in a strict mathematical sense~\cite{Volovik:2003}. Even when quasiparticles behave anomalously compared to conventional metals,   as in one-dimensional electronic systems, this surface is still defined by the same topological invariant. A break up of this surface into hole-like and electron-like pockets, termed reconstruction,  requires a global deformation in the topological sense, most likely  a macroscopic broken symmetry.  

The quantum oscillations referred to earlier are due to the existence of Landau levels in the presence of a magnetic field~\cite{Ziman:1972}, which, in two dimensions,  restructure the energy spectra of the charged quasiparticles  in terms of a discrete set of levels. As the highest occupied level  sweeps past the Fermi energy, as the magnetic field is increased,  the macroscopic quantum state of matter periodically returns to itself, hence the oscillation in a variety of properties. More quantitatively the frequency of oscillation, $F= \frac{\hbar c}{2\pi e} A(E_{F})$,  is the Onsager relation~\cite{Ziman:1972}, where $A(E_{F})$ is the area of a closed orbit in the momentum space at the Fermi energy; here $\hbar$ is Planck's constant divided by $2\pi$, $c$ is the velocity of light, and $e$ is the electronic charge---all fundamental constants. The observed frequency provides a valuable parameter, the Fermi surface area, if the electron orbits are closed, and,  equally importantly, the very existence of the surface itself. There are also surprisingly simple quasiparticles in a superconductor, but  they couple to the magnetic field in such a qualitatively different way that they do not form Landau levels and cannot give rise to the quantum oscillations~\cite{Franz:2000}. In a more technical parlance, their coupling to the gauge field is not a minimal coupling as for normal quasiparticles. Thus, the observation of quantum oscillations reflect normal quasiparticles and their closed orbits on the Fermi surfaces, in particular Fermi pockets. 

The experiments show~\cite{Doiron-Leyraud:2007,LeBoeuf:2007} that in the underdoped  samples of hole-doped cuprates the oscillating $R_{H}$ in high magnetic fields and low temperatures has a {\em negative} sign. Doping refers to adding charge carriers to the  parent compounds of the high temperature superconductors. The superconducting transition temperature is the highest at  a doping that is referred to as optimal. The normal state of less than optimally doped, underdoped, superconductors exhibit a number of anomalous proporties.  That $R_{H}$  oscillates and is negative is striking, because at face value  the robust feature of a given Fermi surface  with conventional quasiparticles, is simply that $R_{H}=\pm 1/nec$, positive for holes and negative for electrons, where $n$ is the density of the charge carriers. How could then the experimental observation be understood? In the present paper we shall attempt to answer this fundamental question and draw its implications. We shall see that an explanation is that the Fermi surface undergoes reconstruction forming both hole and electron pockets and it is these electronic carriers that dominate the sign of the Hall coefficient and the two-pocket picture is the root of the oscillating Hall coefficient.

We propose to explain the quantum oscillation experiments, in particular the Hall measurements,  with the notion of an unusual broken symmetry, namely the $d$-density wave~\cite{Chakravarty:2001} (DDW), which was proposed to unify the phenomenology of high temperature superconductors from a single assumption, but with no special reference to RVB.  In contrast,   the RVB character~\cite{Anderson:1987} was considered essential for its predecessor, the staggered flux state~\cite{Affleck:1988}.  Thus,   one can study DDW  within a suitable effective Hamiltonian even in the weak interaction limit; however, those properties that are determined by the symmetries should be valid even in the strong interaction limit~\cite{Nayak:2000}. We emphasize that it is the notion of a broken symmetry and  a conventional reconstructed Fermi surface that is of fundamental importance in our explanation of these remarkable experiments. 

One of the predictions of DDW was  circulating orbital currents arranged in a staggered pattern that could be detected in neutron scattering experiments because of the resulting magnetic fields, but with very small magnitude of the ordered moments  of the order of $0.05 \mu_{B}$ (Bohr magneton). Such experiments are very demanding and require high quality samples and a proper polarization analysis to detect very small magnetic signals~\cite{Chakravarty:2001b}.  Therefore, attempts to observe DDW  have understandably resulted in controversy. While some experiments purport to confirm it~\cite{Mook:2002,Mook:2004}, others do not~\cite{Stock:2002,Fauque:2006}. From the smallness of ordered moments, one should not immediately infer that all macroscopic signatures of DDW are equally small. A feature of DDW is the  existence of hole and electron pockets in its electronic structure that result in striking consequences, an example of which we shall demonstrate in the present paper.  
At the very least a theory must explain: (a) Why any oscillations of the Hall coefficient  are seen in the first place? (b) How is the Luttinger's sum rule~\cite{Luttinger:1960}  satisfied? (c) Why is the Hall coefficient negative for large fields and low temperatures? The assumption of a {\em static} spin density wave (SDW) can also result in electron and hole pockets~\cite{Chubukov:1997} and can thus lead to a similar explanation of the experiments. However, a critical analysis given below renders this route implausible.

 It is sufficient to consider the behavior at zero temperature, $T=0$, and in the pristine DDW state. At non-zero temperatures, oscillations are expected to be washed out. We assume that the high magnetic field is enough to destroy the DSC component in the coexisting DDW and DSC state in the underdoped regime.  In other words, the DDW state is not a field induced state. We also assume that the lifetime of the DDW quasiparticles at the lowest measured  temperatures are dominated by scattering from impurities and the effects of electronic interactions are negligible in comparison, except insofar as the formation of the ordered state is concerned. It is well known that quantum oscillations are often significantly affected by other complex sources of dephasing, some of which involve sample or magnetic field inhomogenities~\cite{Shoenberg:1984}. These, however, will not be addressed here.

\begin{figure}[htbp]
\begin{center}
\includegraphics[width=\linewidth]{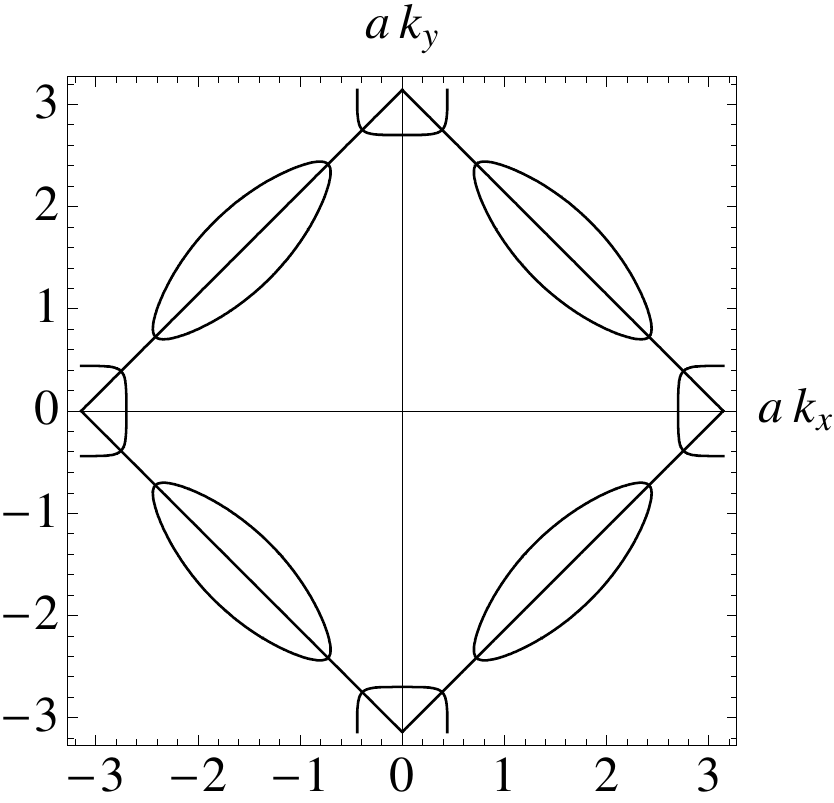}
\caption{The contour plot of the DDW band structure for Y123 with $y=6.51$. There are two inequivalent elliptical hole pockets (one half of four centered at $(\pm\pi/2,\pm \pi/2)$) in the reduced Brillouin zone (RBZ) bounded by $k_{y}\pm k_{x}=\pm \pi/a$ and one electron pocket (one quarter of four centered at $(0,\pm\pi)$ and $(\pm\pi,0)$) defined by the chemical potential $\mu$. A  simplified common set of band structure parameters~\cite{Andersen:1995} are given by $\epsilon_{\mathbf {k}}^{-}= - 2 t (\cos k_{x}a+\cos k_{y}a)$, $\epsilon_{\mathbf {k}}^{+}= 4 t' \cos k_{x}a\cos k_{y}a -2t'' (\cos 2 k_{x}a+\cos 2 k_{y}a)$, where $a$ is the lattice spacing equal to $3.84$ \AA; we ignore the slight orthorhombicity. The parameters are $t=0.3 \; {\mathrm eV}$, $t' = 0.3 t$,  $t''=t'/9.0$, $W(0)= 0.078\;\mathrm{eV} $, and $\mu = -0.2642\; {\mathrm eV}$, corresponding to a total hole doping of $10\%$. }
\label{fig:band}
\end{center}
\end{figure}

First we shall consider commensurate DDW order and then discuss the effect of incommensuration. The commensurate DDW order doubles the unit cell of the real space lattice because the translational symmetry corresponding to a displacement by the lattice spacing $a$ of the square planar CuO-lattice is broken. As a result   the conventional Brillouin zone (BZ) in the reciprocal space is halved, or reduced, which is known as the reduced Brillouin zone (RBZ). This unit cell doubling, but without a  conventional spin or charge density wave order, plays a crucial role in our analysis. 
It is convenient to define
$\epsilon_{\mathbf{k}}^{\pm}= \frac{1}{2}\left[\epsilon_{\mathbf{k}} \pm \epsilon_{\mathbf{k+Q}}\right]$,
where $\epsilon_{\mathbf{k}}$ is the electronic band structure as function of the wave vector $\bf k$, and ${\mathbf Q}$ is the commensurate ordering vector $(\pi/a,\pi/a)$. The elementary excitations, quasiparticles, are given by 
$E_{\bf k}^{\pm}= \epsilon_{\bf k}^{+}
\pm \sqrt{(\epsilon^{-}_{\mathbf{k}})^{2}+W_{\mathbf k}^{2}}$, where 
the DDW order parameter is defined by an energy gap that has the characteristic shape of a $d$-wave, reflecting a bound electron-hole pair:
$W_{\mathbf k} = \frac{W_0(T)}{2}\left(\cos{k_x} a-\cos{k_y} a\right)$. The spectra of excitations in the RBZ consist of two hole pockets and one electron pocket for a generic set of parameters, as shown in Fig.~\ref{fig:band}. 
Despite the appearance of hole pockets, not all experimental probes can detect them. In particular, due to the special coherence factors, only one half of a hole pocket (the inner part) is visible in angle resolved photoemission spectroscopy~\cite{Chakravarty:2003}; as to electron pockets, only the nearly vertical segments close to $(0,\pm\pi/a)$ and the nearly horizontal segments close to $(\pm\pi/a,0)$ should be visible due to the same coherence factor analysis.

Within a conventional Fermi liquid picture and the Boltzmann theory,  Hall resistivity, $R_{xy}$, for a single band, be it a hole or an electron band, is  equal to  $H/(nec)$, with $H$ the magnetic field. Yet, it has been known for sometime that many semimetals with electron and hole pockets exhibit pronounced oscillations of $R_{xy}$ as a function of magnetic field~\cite{Reynolds:1954}, which closely reflect SdH oscillations. It is a textbook exercise~\cite{Ziman:1972} that for two bands the combined Hall coefficient $R$ is given by (for magnetic fields of interest)
\begin{equation}
R\approx\frac{R_{1}\sigma_{1}^{2}+R_{2}\sigma_{2}^{2}}{(\sigma_{1}+\sigma_{2})^{2}}
\label{eq:Hall}
\end{equation}
where $R_{i}$  and $\sigma_{i}$ are the individual  Hall coefficients and the conductivity of the $i$th band. The formula makes it transparent that  the oscillations of the conductivities can lead to oscillations of $R$, and its sign depends on the relative contributions of the bands 1 and 2, respectively a hole and an electron band. There are a number of implicit assumptions: that independent Hartree-Fock quasiparticle approximation is valid, the Boltzmann theory of transport is adequate, and there is no quantum tunneling between the bands. The first two assumptions are the {\em sine qua non} of the DDW theory, in which, well inside the ordered phase, the order is described by mean field theory and its elementary excitations are simple quasiparticles.  The last assumption can be correct only if the quasiparticle gap is not so small that the quantum tunneling effects dominate and result in a reconnection of the Fermi surfaces. 

Ando ~\cite{Ando:1974}  has derived a marvelous  formula for $\sigma$ for a non-interacting continuum two dimensional electron system, but including impurity scattering within a self-consistent Born approximation, in a perpendicular magnetic field and Fermi energy situated at a high Landau level. In general, the Landau level problem in a crystal is not so simple, but in the limit of high Landau level, $N$,  and in the absence of interband transitions, both of which are excellent assumptions here, we can apply the  quantization condition for a closed orbit:
$
\oint k_{x} dk_{y} = \frac{2\pi e H}{\hbar c} \left[N+\gamma(N)\right],
$
where $\gamma(N)$ is a number between 0 and 1. The relevant chemical potential, $\mu$, which is also $E_{F}$ at $T=0$, is so large  that the Dirac character of the nodal fermions is simply irrelevant. Ando's result for the continuum can be recast in a slightly more general form by using the above quantization condition, which becomes 
\begin{equation}
\sigma =\sigma_{0}\frac{1}{1+\phi^{2}}\left[1+4\frac{\phi^{2}}{1+\phi^{2}}\sum_{s=1}^{\infty}e^{-s\pi/\phi}\cos\left(s\frac{2\pi F}{H}\right)\right],
\label{eq:conductivity}
\end{equation}
where $\phi=\omega_{c}\tau =\frac{eH}{m^{*}c}\tau\lesssim 1$, $m^{*}$ is the effective mass, and $\tau$ is the elastic scattering time from the impurities. Because the higher harmonics fall off rapidly, because of the exponential damping factors, called the Dingle factors, here we will consider only $s=1$. 
Note that  the Drude conductivity $\sigma_{0}=ne^{2}\tau/m^{*}$  is in general  different for different bands. Similarly, $\phi$ may be different for bands that are distinct.   Here $\tau$ is the impurity scattering time, but it is quite possible that additional dephasing must be included if the sample or the magnetic field is inhomogeneous. Note that in applying Eq.~\ref{eq:conductivity} to Eq.~\ref{eq:Hall}, the absolute magnitudes of the individual $\sigma$'s do not enter, but only their ratio. It is also very reasonable that the ratio of $\phi$'s should reflect the ratio of $\sigma$'s.

\begin{figure}[htbp]
\begin{center}
\includegraphics[width=\linewidth]{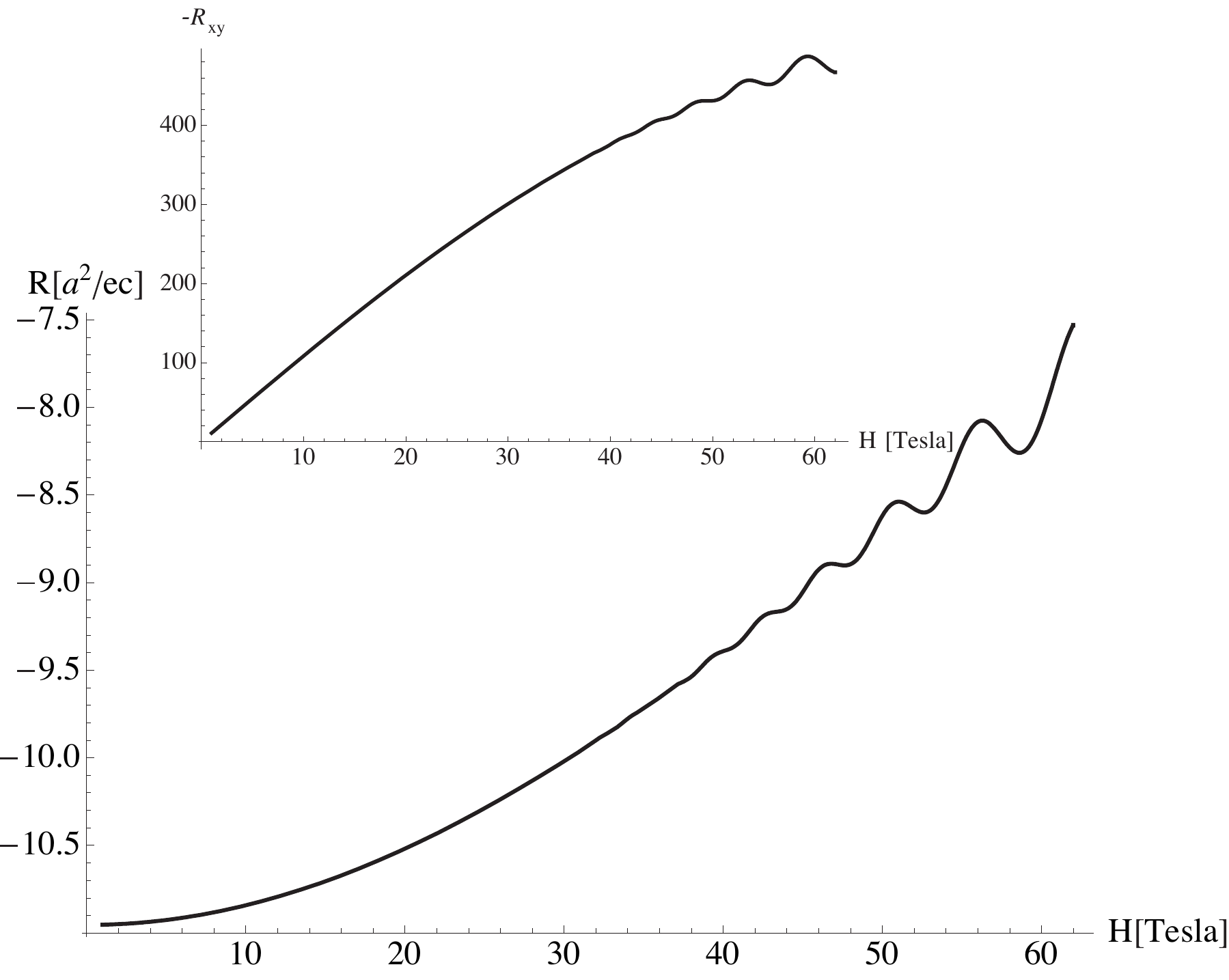}
\caption{Oscillations of the Hall coefficient $R$ in $\mathrm{YBa_{2}Cu_{3}O_{y}}$ with $y=6.51$; the unit corresponds  two dimensional areal density. The relevant band structure parameters are the same as in Fig.~\ref{fig:band}. The additional parameters needed are $\sigma_{0}^{(1)}/\sigma_{0}^{(2)} = 1/2$, $\phi_{1}/H=0.0065\; \mathrm{T^{-1}}$ and $ \phi_{2}/H=0.013\; \mathrm{T^{-1}}$, as discussed in the text. The oscillation period $ \mathrm{1/530 T}$ is entirely determined by the area of the electron pocket. The oscillations from the hole pocket are not visible because of the exponentially small Dingle factor. The inset is the negative of the  Hall resistivity $-R_{xy}=-R H$, where $H$ is expressed in terms of Tesla.}
\label{fig:RHall}
\end{center}
\end{figure}

It is now straightforward to plot $R(H)$ using Eq.~\ref{eq:Hall}. The results for a set of parameters are shown in Fig.~\ref{fig:RHall}. To obtain a single oscillation frequency of $R$, as observed in experiments and with a negative sign, the oscillations must arise from the electron pocket; the constraint of the Luttinger sum rule~\cite{Luttinger:1960} as modified to apply to the RBZ~\cite{Chubukov:1997} provides a very severe constraint. The most general form of this sum rule, also a rigorous theorem, states that the particle density is twice (for two spin directions)  the ``volume'' of the momentum space in $d$-dimensions  divided by $(2\pi\hbar)^{d}$ over which the real part of the single particle Green function at the Fermi energy is positive. 

The principal results are relatively insensitive to the band structure parameters as long as this sum rule is satisfied and the electron and hole pockets have the right curvature. Thus we have chosen parameters such that over the range of the magnetic field the contribution from the hole pocket is smooth and does not vary much. This becomes obvious if one plots separately the individual conductivities $\sigma_{1}$ and $\sigma_{2}$. Of course, this is true only  for the range of parameters used  in our calculation, but with a different set of parameters one can observe oscillations arising from the hole pockets as well, and in general one can observe complicated oscillatory patterns.

It is important to emphasize the details of our choice of parameters to expose the robust features outlined above.  We have employed a simplified version of the band parameters given in Ref.~\cite{Andersen:1995} for $\mathrm{YBa_{2}Cu_{3}O_{y}}$ with $y=6.51$; see the caption of Fig.~\ref{fig:band}.  We have verified that other choices of parameters lead to similar results.
Once these parameters are given and the doping level is fixed, there is no further flexibility in choosing $F_{i}$ or $R_{i}$. The $F_{i}$'s are determined by the area of the Fermi pockets. The corresponding densities, hence $R_{i}$, are obtained by applying Luttinger's counting argument to the RBZ. Including  a factor of 2 for spin and a factor of 2 for the two hole pockets, the density corresponding to the hole pockets, $n_{1}\equiv n_{h}$, is $n_{h}= A_{h}/\pi^{2}$, where $A_{h}$ is the $k$-space area of a single hole pocket.  We convert $n_{h}$ to number of carriers per Cu-atom by multiplying it by $a^{2}$, but maintain the same notation, as no confusion should arise. For the density in the electron pocket we have to include again a factor of 2 for spin but there is now only one electron pocket in the RBZ, that is, $n_{2}\equiv n_{e}=A_{e}/2\pi^{2}$. Now $R_{1}=1/n_{h}ec$ and $R_{2}=-1/n_{e}e c$, where the charge $e > 0$.

The remaining parameters are $\sigma_{0}$ and $\omega_{c}\tau$ for the two bands. As mentioned above, Eq.~\ref{eq:Hall} is independent of the absolute value of $\sigma_{0}$; it is only the ratio between the two bands that matter. In the present set of experiments there is evidence for oscillations involving only a single frequency: $\Delta(1/H)=1/530\mathrm{T}$. The single frequency implies that only one of  the bands dominates the oscillation,
and its period determines  the area of the relevant Fermi pocket via the Onsager relation.
To understand which band is responsible for the oscillation, let us first assume that
it is the hole pocket. Note that the hole doping, $\delta$, for $\mathrm{YBa_{2}Cu_{3}O_{6.51}}$ is  0.1. However, the density corresponding to two hole pockets  in the RBZ,
deduced from the oscillation period, is $2\times 0.038=0.076$, which leads
to  a violation of the Luttinger sum rule independently  of any electron pockets.
Taking into account an electron pocket will result in a $\delta$ less than $0.076$, an even greater violation of the sum rule, because $\delta = n_h - n_e$. If, instead, the oscillations are from the electron pocket,  $F_2$ is $530$ T,  and therefore  $n_e =0.038$.
To achieve a  hole doping of 0.1, we must set $n_h =0.138$, implying $F_1 =969$ T. Therefore,  the single observed frequency of 530 T combined with the Luttinger sum rule uniquely constrain $n_{e}$, $n_{h}$, $F_{1}$ and $F_{2}$.  Moreover, because only one frequency is observed we must have  $\omega_c \tau$ for the electron
pocket larger than $\omega_c \tau$ for the hole pocket. The Dingle factors are critical for this and the final parameters are given in the captions of Fig.~\ref{fig:band} and  Fig.~\ref{fig:RHall}. 

The experiments observe the evolution of the superconducting state to the normal state as the magnetic field is increased up to 62 T. Clearly in the superconducting state both $R_{xx}$ and $R_{xy}$ are zero, but in the vortex liquid state these quantities are finite.  It is claimed~\cite{LeBoeuf:2007}, however, that for $H \gtrsim 45\; \mathrm{T}$ the behavior truly reflects the normal state, free of traces of superconductivity, and it is in this experimental regime that our theory is relevant. Overall, we find the agreement with experiments excellent.   We have focused here on oscillations of $R$ because it is the most striking consequence of the existence of both electron and hole pockets,  but clearly there will be also SdH and dHvA oscillations with the same frequency $F_{2}$. 

We wish to emphasize that our approach is robust and equally well describe the results for Y124. Although Y123 and Y124 must necessarily have different band structures because they are different materials, the required adjustment is truly minimal. We can fit the results for Y124 with all parameters the same as in  Fig.~\ref{fig:band} except for   $t''=t'/5.5$, $W(0)= 0.051\;\mathrm{eV} $, and $\mu = -0.286\; {\mathrm eV}$, corresponding to a total hole doping of $12.5\%$.  Once again, the Luttinger sum rule, the value of doping, and the observed areas of the quantum oscillations almost uniquely fix all the parameters. The additional parameters needed are the same as in Fig.~\ref{fig:RHall}. The oscillation frequeny $ \mathrm{660 T}$ is entirely determined by the area of the electron pocket. Note that the required value of the DDW gap $W(0)$ is now decreased to $\mathrm{0.051 eV}$, which is consistent with the DDW theory, because the DDW gap should decrease with increased doping~\cite{Chakravarty:2001}.

Until recently~\cite{Doiron-Leyraud:2007,LeBoeuf:2007} no signatures of electron pockets have been detected in hole doped cuprates, nor are they predicted by the electronic structure calculations based on local density approximation in either Y123  or Y124; see Refs.~\cite{Doiron-Leyraud:2007,Bangura:2008} and references therein. 

Previously, electron pockets were noted in the DDW band structure but their presence~\cite{Chakravarty:2002} did not make any critical difference. The aim there was to explain the nonanalyticity in the Hall number as the DDW gap collapses in the middle of the superconducting dome at a quantum critical point. As mentioned above, a static commensurate SDW  can also lead to both hole and electron pockets~\cite{Chubukov:1997,Lin:2005}, but  this  is not experimentally relevant in the  doping regime of interest here. Moreover, conventional SDW order parameter is necessarily a $s$-wave object and clearly inconsistent with the strongly momentum dependent $d$-wave character of the pseudogap~\cite{Norman:1998}. A triplet version of DDW which results in a staggered circulating spin currents does have the necessary $d$-wave character and the associated electron pockets, but is implausible for other reasons.\cite{Chakravarty:2001} It has been recently argued that a high magnetic field can induce a conventional SDW order and the experiments have been interpreted in terms of two hole pockets~\cite{Chen:2007}.  But this approach clearly cannot explain the negative Hall coefficient and its oscillations, nor the Luttinger sum rule.  The low energy spin fluctuations in YBCO typically have a spin gap below the $(\pi/a,\pi/a)$ resonance peak at higher doping, which seems not to be the case for $\mathrm{YBa_{2}Cu_{3}O_{6.5}}$, where the spectral weight in the normal state vanishes linearly with energy~\cite{Stock:2004}. However, it is difficult to see, simply from energetics,  how even a 65 T magnetic field can produce a condensation at zero energy of a static SDW  with sufficient spectral weight to produce a large enough order parameter that is necessary to obtain the required sizes of the pockets. An interesting discussion along similar lines of the implausibility of SDW or antiferromagnetic order determining these experiments was also offered in Ref.~\cite{Lee:2007}.

In a mean field theory it was found~\cite{Millis:2007} that $1/8$ magnetic antiphase stripe order generically produces complicated Fermi surfaces involving open orbits, hole pockets and electron pockets. However, it appears to be difficult  to satisfy simultaneously the constraints of
the Luttinger sum rule,  the periodicity of the oscillations, and the negative sign of $R$, consistent with the observations  in Refs.~\cite{Doiron-Leyraud:2007,Bangura:2008,LeBoeuf:2007}. 
A more exotic mechanism has also been invoked to explain these experiments~\cite{Kaul:2007}, but within this mechanism the correct negative sign of the Hall coefficient is very difficult to achieve. 

What, we may ask, will be the scenario if the order, in particular the DDW order, incommensurates, which is a distinct possibility over a range of hole doping?~\cite{Kee:2002} Although it is energetically expedient to add holes at the nodes, unlike SDW, the Fermi surface must eventually move away from the nesting wave vector as a function of doping. A rigorous treatment of any incommensurate order is generally complex, especially if the incommensuration is irrational. Nonetheless, the dominant physical behavior can be often approximated by the lowest order gap and the hierarchy of gaps will be washed out due to thermal fluctuations, disorder, or magnetic breakdown~\cite{Falicov:1967}. Thus, in a simplified treatment, the DDW order parameter can be crudely approximated by the quantum mechanical expectation value~\cite{Dimov:2005},
\begin{equation}
\langle c^{\dagger}_{\sigma'\mathbf k'} c_{\sigma\mathbf k}\rangle = i \frac{W_{\mathbf k}}{2} \left(\delta_{\mathbf {k', k+K}}+\delta_{\mathbf {k', k-K}}\right) \delta_{\sigma,\sigma'},
\end{equation}
where the incommensuration vector ${\mathbf q}= {\mathbf K}-(\pi/a,\pi/a)$; $\sigma$, $\sigma'$ are spin indices and $c_{\mathbf k}$, $c_{\mathbf k'}^{\dagger}$ are the Fermion destruction and creation operators. This {\em Ansatz} conserves current to only linear order in $\mathbf q$, which may be sufficient for practical situations. 

In high temperature superconductors, there is some evidence from neutron scattering of incommensurate SDW fluctuations that the incommensuration is of the form ${\mathbf q}= \pi(\pm  2\eta,0)/a$ and ${\mathbf q}= \pi(0,\pm  2\eta)/a$, where $\eta \sim 0.1$ in the underdoped YBCO~\cite{Dai:2001}. We believe that similar estimates should also apply for incommensurate DDW because the determining competition between the kinetic and the interaction energies are similar within a mean field theory; the precise value of $\eta$ is not particularly relevant at the level of present discussion. If we further simplify by assuming a single wave vector $\mathbf q$, by spontaneous breaking of inversion symmetry, the excitation spectrum can be trivially solved. One can easily show, using the parameters essentially identical to those as before,  that for 10\% doping  the frequency of one of the hole pockets can be shifted from 969 T to a higher frequency  with little  change in the frequency of the electron pocket ($530$ T), but generically there is an additional smaller hole pocket of lower frequency, tightly constrained by the Luttinger sum rule. The physics of the Fermi surface reconstruction is completely unchanged as compared to the commensurate case. The same assumption that the hole pockets have lower mobility will render their observation difficult, but still the splitting of the frequencies of the hole pockets is a prediction of the incommensurate picture. In addition, the field sweep necessary for the observation of the smaller hole pocket corresponding to a lower frequency may be a limiting factor in unambiguously identifying it without higher field magnets. 

We have assumed, along with Ref.~\cite{LeBoeuf:2007}, that, once there is
no longer any non-linear field-dependence to the transport,  the system is in the  ``normal state'' , that is, there are no  contributions from vortices. One can
go even  further and claim that the samples are above the upper critical field, $H_{c2}$, and that there are
no vortices around at all. This need not be the case. It is quite possible  that these high field experiments are still 
well below $H_{c2}$, guessed to be 100T or more.  It is well known and understood, however,  that quantum oscillations in many superconductors are observed to fields as small as $(1/2) H_{c2}$, with the oscillation frequencies unchanged from the non-superconducting or the ``normal state'', but with an increased damping~\cite{Wasserman:1996}. As remarked above, it is also known that the quasiparticles of the high temperature superconductors do not form Landau levels in the superconducting state~\cite{Franz:2000}. Thus, the very fact that the oscillations are observed in three different samples, with three independent techniques, and by at least two independent groups imply that the  experiments are accessing the  normal state beyond the realm of superconductivity. A convincing theory of the amplitude of the oscillations is a very difficult problem, however.

The seeming lack of observation, yet, of electron and hole pockets in other measurements in hole-doped superconductors, in particular in angle resolved photoemission spectroscopy that is also capable of measuring the Fermi surface is an important puzzle; see, however, the work on electron-doped materials~\cite{Armitage:2001} where both electron and hole pockets are observed. Clearly further experiments are necessary to settle these important issues more definitively, but our simple symmetry  breaking approach remains a serious challenge to alternative scenarios, which has to be consistent with all the experimental facts, in particular the negative Hall coefficient, as well as the theoretical constraints of the Luttinger sum rule, known to be valid for a whole class of systems, including Mott insulators~\cite{Dzyaloshinskii:2003}. There is another serious implication of our work. That a simple quasiparticle/Fermi surface based theory can explain these striking data in the underdoped regime, which previously was thought to be plagued by the complexities ensuing from the proximity of a Mott insulator, is a warning that we may have misidentified the effective Hamiltonian as the single band Hubbard model. In particular, the correct effective Hamiltonian should include correlated hopping processes that are necessary to stabilize both DSC and DDW~\cite{Nayak:2000}. The properties protected by symmetries can equally well be understood in the weak interaction limit with a proper effective Hamiltonian.

\begin{acknowledgments}
This work was supported by the National Science Foundation, Grant. No. DMR-0705092 (S. C.) and by the Canadian Institute for Advanced Research, the Canada Research Chair, and the NSERC of Canada (H. Y. K.). We would like to thank the Aspen Center for Physics and H. Dennis Drew and P. A. Lee for important comments. The consideration of incommensurate DDW was motivated by the recent unpublished work of S. E. Sebastian and her collaborators (S. E. Sebastian, N. Harrison, and G. Lonzarich, private communication).
\end{acknowledgments}

\end{article}

\end{document}